# Multi-modal machine learning analysis of GaSe molecular beam epitaxy growth conditions

*Mingyu Yu,* [†] *Isaiah A. Moses,* [‡] *Wesley F. Reinhart,* [§, ‖] *Stephanie Law*[*,§]

[†]Department of Materials Science and Engineering, University of Delaware, Newark, Delaware 19716, United States

[‡]Materials Research Institute, Pennsylvania State University, University Park, Pennsylvania 16827, United States

[§]Department of Materials Science and Engineering, Pennsylvania State University, University Park, Pennsylvania 16827, United States

[‖]Institute for Computational and Data Science, Pennsylvania State University, University Park, Pennsylvania 16827, United States

[*]Corresponding author: Stephanie Law

Email: slaw@psu.edu

ABSTRACT: Autonomous synthesis platforms integrating machine learning with *in-situ* diagnostics have the potential to revolutionize thin-film growth by enabling real-time process optimization and reducing the need for manual tuning. However, their application to molecular beam epitaxy (MBE) remains underdeveloped. Here, we present a machine learning-guided framework for MBE growth of GaSe films, leveraging reflection high-energy electron diffraction (RHEED) as an *in-situ* diagnostic alongside *ex-situ* characterization via X-ray diffraction and atomic force microscopy. Unsupervised learning on RHEED patterns reveals a well-defined boundary between high- and low-quality samples, capturing physically meaningful features. Mutual information analysis shows a strong correlation between RHEED embeddings and rocking curve full-width at half maximum (FWHM), while the correlation with AFM root-mean-square (RMS) roughness is weak. Among key growth conditions, growth rate most strongly influences FWHM, whereas the Se:Ga flux ratio primarily affects RMS roughness and the RHEED embeddings. Supervised learning models trained to predict FWHM and RMS roughness demonstrate moderate accuracy, with significant improvement achieved by incorporating RHEED embeddings. Furthermore, anomaly detection *via* residual analysis in supervised learning aligns well with unsupervised classification from RHEED, reinforcing the reliability of the predictive models. This study establishes a data-driven framework for machine learning-assisted MBE, paving the way for real-time process control and accelerated optimization of thin-film synthesis.

## 1. INTRODUCTION

The rapid advancements in machine learning, artificial intelligence, and high-performance computing are providing transformative technological tools for the exploration and synthesis of new materials.[1-5] Machine learning models, constructed by integrating theoretical and empirical approaches, hold the potential to explore parameter space autonomously, quickly establish process-performance relationships, and diagnose material synthesis in real time.[6,7] This reduces reliance on manual intervention in parameter space exploration, enabling more precise and efficient mechanistic control — an essential direction for advancing materials science.[8,9]

Typically, material synthesis workflows involve characterization techniques, such as X-ray diffraction (XRD), atomic force microscopy (AFM), and photoluminescence, to analyze and optimize material properties. Solution-based synthesis methods can already be integrated with these characterization tools to form commercial pipeline systems.[1,10,11] In contrast, vapor phase deposition methods have yet to achieve efficient autonomous production platforms.[12-14] For molecular beam epitaxy (MBE), despite its breakthroughs in synthesizing materials such as chalcogenides, III-V compounds, oxides, and nitrides — particularly its unparalleled ability to produce high-quality, atomically thin crystalline films — its stringent growth conditions and complex epitaxial mechanisms make the process of optimizing growth process time-consuming and expensive.[15-20] Therefore, leveraging machine learning to develop autonomous MBE growth platforms presents a highly promising prospect. A Bayesian approach was reported to optimize metal-organic MBE synthesis for highly crystalline TiN films.[21] C. C. Price et al.[22] developed predictive models using *in situ* reflection high-energy electron diffraction (RHEED) data for *ex situ* analysis. However, research on employing machine learning for real-time anomaly detection remains limited.

Our study on the multi-modal machine learning-guided MBE synthesis is based on a comprehensive high-quality dataset of GaSe thin films grown on GaAs (111)B substrates.[23,24] GaSe is an emerging two-dimensional (2D) semiconductor material with intriguing properties, including thickness-tunable bandgaps, nonlinear optical behaviors, and intrinsic p-type conductivity,[25-31] making it a promising candidate for applications in electronics, optics, and optoelectronics.[32-37] Moreover, as a representative member of the van der Waals (vdW) chalcogenide semiconductor family,[38-40] insights gained from studying GaSe can be extended to other vdW chalcogenides. The ultra-high vacuum environment of MBE, along with precise control over source materials and temperature, enables the synthesis of GaSe thin films with high purity, well-ordered crystallinity, controlled layer thickness, and smooth surfaces.[23,24,30,41-47] However, optimizing the interrelated growth conditions experimentally necessitates extensive trial and error.

In this work, we aim to leverage machine learning to analyze the relationships between various growth conditions (Ga flux, which represents growth rate as Ga is the limiting element in GaSe MBE growth, Se:Ga flux ratio, and substrate temperature) and the resulting sample quality, as well as the correlations among various characterization results. This approach seeks to predict sample properties based on input growth conditions and autonomously adjust the synthesis process in real-time using feedback from *in-situ* diagnostic technique —RHEED, which reveals real-time information on crystallinity and surface quality of top layers during MBE growth.[48-51] The main *ex-situ* characterization modalities include high-resolution XRD ω scan — full-width at half maximum (FWHM) of rocking curves are used to quantify the crystallinity of the film; and AFM — providing root-mean-square (RMS) roughness of the surface. Previous studies[52-55] have reported the application of principal component analysis (PCA) and $k$-means clustering methods to identify features and trends in RHEED signals and have applied these approaches to the growth of perovskite, $Fe_xO_y$, and transition metal dichalcogenide thin films. Our unsupervised learning model is capable of autonomously interpreting the physical meaning embedded in RHEED patterns and identifying the crystallinity of the samples. In addition, machine learning has been reported to effectively classify AFM images and recognize surface features,[56-58] as well as extract information related to phase, lattice, defects, and substituents from XRD measurements.[59]

Our research focuses on understanding the relationships between growth conditions and these characterization results, as well as the correlation between *in situ* and *ex situ* characterization metrics, aiming to achieve quantitative analysis through machine learning. Mutual information (MI) analysis identified the growth conditions with the most significant impacts on crystallinity and surface roughness, as well as the correlations among RHEED patterns, FWHM crystallinity, and RMS roughness. These findings were further validated by feature importance analysis. Supervised learning, on the other hand, constructed predictive models for two characterization quantities, RMS roughness and FWHM crystallinity, both achieving moderate predictive accuracy. Notably, the combination of the RHEED embeddings and the FWHM prediction model not only enhanced FWHM predictability but also demonstrated the capability of RHEED embeddings to detect anomalous samples. Our work provides valuable insights for developing machine learning-guided experimental design and contributes to unlocking the full potential of RHEED as an *in-situ* monitoring tool.

## 2. EXPERIMENTAL SECTION

**2.1. MBE Synthesis.** GaSe film samples in this study were grown on epi-ready GaAs(111)B substrates using a DCA Instrument R450 MBE reactor (instrument details at DOI:10.60551/gqq8-yj90). The GaAs (111)B wafers, ordered from Wafer Tech, have a primary flat oriented along the $[01\bar{1}]$ direction. Before use, wafers were diced into 1cm × 1cm pieces and ultrasonically cleaned sequentially in acetone, isopropanol, and deionized water for 10 minutes each. After drying with nitrogen, the substrates were immediately transferred to the load lock chamber and degassed at 200 ºC in a vacuum of about $5 \times 10^{-7}$ Torr for 2 hours to remove residual contaminants. Prior to GaSe growth, the substrate was deoxidized and Se-passivated in the main chamber, as detailed in Ref. 23, and then cooled to the designated growth temperature. Ga and Se fluxes were independently supplied by Knudsen effusion cells, which were measured at the substrate position using a quartz crystal microbalance. The growth temperature was monitored with a non-contact thermocouple positioned behind the substrate. As thermocouple readings typically deviate from the actual substrate temperature, we provide a reference point: in our MBE system, the thermocouple measured 680 °C for GaAs substrate deoxidation,[23,24] which typically occurs at a temperature of about 580 °C.[60,61] All GaSe films in this study were grown to a target thickness of 24 nm.

RHEED was used for real-time monitoring of the sample during growth. It is a powerful tool commonly employed in MBE to observe the crystallinity and morphology of the growing surface. Generally, for smooth single-crystal epitaxial films, incident electrons diffract from the surface layers, producing streaky patterns on the RHEED screen as the Ewald sphere intersects the reciprocal lattice rods perpendicular to the surface. For rough single-crystal films with islands, electrons transmit through the islands and diffract from near-surface planes, forming sharp spot patterns. Polycrystalline films generate continuous ring patterns due to electron diffraction from randomly oriented planes, while amorphous surfaces exhibit no discernible diffraction patterns.[62] RHEED patterns can also reveal twin defects and stacking faults, as streak spacing and arrangement are closely linked to the surface lattice structure and reconstruction.[63,64] The RHEED patterns used in this study were taken at the end of the growth of each sample.

This study included 37 GaSe samples in the machine learning analysis. We used a Sobol sequence to generate growth parameters for the 37 samples. Detailed growth parameters and sample IDs are provided in Table S1, indicating a wide distribution in the parameter space of Ga flux, Se:Ga flux ratio, and growth temperature.

**2.2. *Ex-situ* Characterizations.** High-resolution XRD measurements were conducted on a Malvern PANalytical 4-Circle X'Pert 3 diffractometer with a Cu-K$\alpha_1$ source. ω scans were rocked around the GaSe (004) diffraction peak, the most intense characteristic peak of GaSe crystals. The ω scan range was 7.113 ° to 15.103 °, with a step size of 0.01 ° and a counting time of 0.44 seconds per step. The FWHM of the rocking curves quantifies the crystallinity of GaSe films, where a larger FWHM indicates a higher defect density. These values were extracted from the raw data using the X-ray Utilities Python package. Surface morphology and RMS roughness were examined using a Bruker Dimension Icon AFM. Measurements were taken over a 1 μm × 1 μm area with a resolution of 512 × 512 pixels and a scan rate of 1 Hz. The RMS roughness was determined as the root-mean-squared error between the Gaussian smoothed and the original image. Table S2 summarizes the FWHM and RMS roughness data for all GaSe samples, as well as the RHEED embedding components $Z$. All *ex-situ* characterizations were completed within 24 hours of sample removal from the MBE chamber to minimize GaSe degradation. The raw data for synthesis and characterization can be found in Ref. 65.

## 3. MACHINE LEARNING

**3.1. Unsupervised Learning.** We applied the Uniform Manifold Approximation and Projection (UMAP) method for unsupervised representation learning (implemented in the umap-learn Python package).[66] UMAP constructs a nearest-neighbor graph to generate a low-dimensional manifold with approximately constant density while preserving the topology of the original feature space.

UMAP accounts for non-linear relationships in the data, unlike other low dimensional representations such as the principal component analysis (PCA)[67] and non-negative matrix factorization (NMF).[68] While clustering approaches might also be suitable for identifying groups of similar samples, embedding offers the possibility of multiple feature axes and completely smooth class boundaries. Furthermore, since we do not know *a priori* whether discrete groups should exist in the dataset, dimensionality reduction is more appropriate.

None of these common techniques, including UMAP, directly lend themselves to embedding image data, as pixel-wise lightness values suffer from spurious translation, rotation, and scale dependencies. Therefore, we designed a custom distance metric to compare RHEED samples as follows:

$$d_{\mathrm{RHEED}}(R_i, R_j) = \min_{\theta_i,\theta_j} \left| W(R_i, \theta_i) - W(R_j, \theta_j) \right|_2 \quad (1)$$

where $R_i$ indicates RHEED image $i$, $W$ indicates a $256 \times 512$ – pixel window taken from $R_i$ as parameterized by $\theta_i$, and $|\cdot|_2$ indicates the $L_2$ norm. The parameterization $\theta_i = (x_i, y_i, \alpha_i)$ determines the center $(x_i, y_i)$ of the window $W$ within $R$ while $\alpha_i$ determines the rotation angle of $W$ with respect to $R$. Images were used directly as captured from the RHEED instrument. The minimization was performed with the scipy implementation of the Nelder-Mead simplex algorithm. The final distance metric used in the UMAP embedding is the minimum over all RHEED taken on a given sample:

$$d_{\mathrm{sample}}(\{R\}_i, \{R\}_j) = \min_{k,l} d_{\mathrm{RHEED}}(R_k, R_l) \quad (2)$$

where $\{R\}_i$ indicates the complete set of RHEED images taken for a sample $i$, such that the final distance $d_{\mathrm{sample}}$ is defined once per pair of samples by the best-aligned pair of RHEED images from the two samples. Practically, this avoids the ambiguity of trying to decide which plane is represented by amorphous samples; both GaSe $[1\bar{1}00]$ and $[11\bar{2}0]$ planes were measured from highly crystalline samples, but only one measurement was taken from weakly crystalline samples due to their similar appearance.

**3.2. Mutual Information (MI).** MI quantifies how much knowing one variable's value reduces uncertainty about another variable.[69] We estimated this by examining the local density of data points in both the joint space $(X, Y)$ and the individual spaces (i.e., $X$ and $Y$ separately). When $X$ and $Y$ are strongly related, points that are close together in $X$ tend to be close together in $Y$ as well, resulting in higher MI values. Conversely, when they are independent, knowing a point's location in $X$ provides no information about its location in $Y$, resulting in MI close to zero. Like UMAP, MI accounts for non-linear trends within the data. We estimated MI using a $k$-nearest neighbor approach based on the Kraskov-Stögbauer-Grassberger (KSG) algorithm,[70] given by the function:

$$MI(X, Y) = \psi(n) + \psi(k) - \langle \psi(n_x + 1) + \psi(n_y + 1) \rangle \tag{3}$$

where $\psi$ is the digamma function (for normalization), $n$ is the sample size, and $k$ is the number of nearest neighbors (we used $k = 3$). $n_x$ and $n_y$ are the numbers of points within a given radius in the marginal spaces $X$ and $Y$, respectively. The algorithm used the Chebyshev distance metric for nearest-neighbor searches. Data were preprocessed using the Standard scaler in scikit-learn. The MI implementation was also adapted from the scikit-learn library's mutual information estimator.

**3.3. Supervised Learning.** It was performed to model the relationship between MBE growth recipe and resulting measurements. The small number of available samples makes deep learning unsuitable. Therefore, we used a Random Forest model[71] due to the nonlinear nature of the task and the superiority of decision tree ensembles for modeling tabular data. Random Forest also provides an advantage in that no preprocessing is required to handle variable distribution or magnitude; all inputs were provided in the form indicated in Tables S1 and S2. While some hyperparameter tuning was investigated, no compelling relationship between hyperparameters and model validation performance was found. As a result, the default model hyperparameters for the scikit-learn implementation were used (most importantly, 100 estimators and no maximum tree depth), except for specifying a fixed random state. We also note that other standard regression models available in scikit-learn were investigated, such as linear regression (and regularized variants), polynomial regression, Gaussian Process, gradient boosting, and shallow neural networks. None of these gave improved performance, and many required more complex feature engineering (such as rescaling), making Random Forest the most appealing choice.

Due to the limited data, the regression models were trained and evaluated using Leave-One-Out (LOO) cross-validation (CV),[72] so for $N$ samples, $N$ models were trained on the remaining $N-1$ observations and evaluated against the one held-out observation. This emulates deploying the model to grow one new sample $N$ times, once for each sample we have, and scores the result. This method was selected due to the small size of the available dataset and high variability between samples, which made typical cross-fold validation difficult to interpret. Regressors were evaluated on the held-out data by computing the coefficient of determination, $R^2$, over the entire held-out set. Reported $R^2$ values thus reflect the ensemble of $N$ regression models trained on $N$ different folds, more generally describing a modeling strategy rather than a specific trained model. The machine learning processed data and code can be found in Ref. 73.

## 4. RESULTS AND DISCUSSION

**4.1. Unsupervised RHEED Embeddings.** We began with unsupervised learning on RHEED patterns as they provided real-time information. Figure 1 shows a UMAP projection of RHEED pattern, where closer data points indicate greater similarity in RHEED signals, illustrating how the RHEED patterns vary with the $Z_0$ and $Z_1$ embeddings. As RHEED serves as a qualitative indicator of the growth surface condition: sharp, well-defined, and periodically occurring streaky patterns typically signify high crystallinity and smoothness; the appearance of dot or dashed patterns indicates the formation of protrusion features, while blurred, broadened streaks suggest a decline in crystallinity. Figure 1 shows a smooth transition at $Z_0 = 1$ between the two regions. Samples with $Z_0 > 1$ are labeled as "low quality", while those with $Z_0 < 1$ are considered "high quality". The low-quality samples exhibit spotty RHEED patterns associated with 3D island formation. In contrast, high-quality samples generally display sharp, continuous streaks, indicative of an atomically smooth crystalline surface. Within this set, $Z_1$ generally correlates with surface crystallinity, where lower values correspond to sharper streaks (higher crystallinity). However, a few high-quality samples, such as Samples #7 and #33, exhibit weak and broad streaks, indicating reduced crystallinity. This suggests that the spatial arrangement is not solely based on sample quality. As discussed in Section 4.3.3 on anomaly detection, the so-called "low-quality" samples are more likely those that deviate from normal growth behavior, thus compromising modeling accuracy and reliability. These findings suggest that the latent embedding derived from unsupervised learning on RHEED data effectively captures physically meaningful features of the images and reveals certain trends, providing a robust foundation for further analysis.

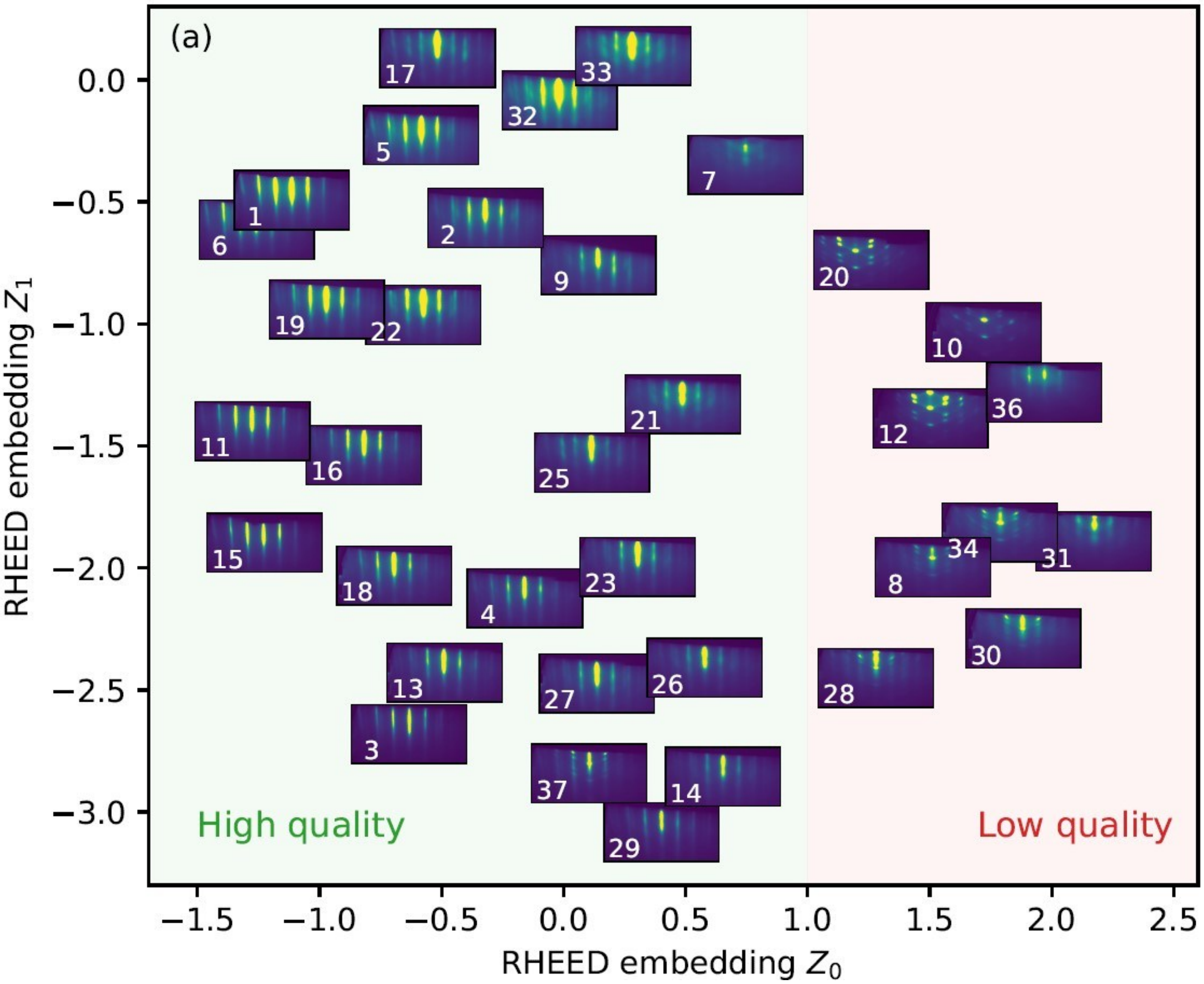

**Figure 1.** Unsupervised UMAP embedding of RHEED signals. The spatial arrangement of RHEED signals in the figure illustrates how they vary over the $Z_0$ and $Z_1$ embeddings. Sample numbers are indicated in the lower left of each RHEED panel. The RHEED embeddings $Z_0$ and $Z_1$ have no physical units. “High quality” and “low quality” refer to samples that follow and deviate from normal growth behavior, respectively, as discussed in Section 4.3.3.

**4.2. Mutual Information (MI) Analysis.** Next, we computed the MI between individual growth conditions (Ga flux, Se:Ga flux ratio, and growth temperature) and various characterization parameters, including the learned RHEED embedding $Z$ and conventional scalar response variables such as rocking curve FWHM and AFM RMS roughness. This analysis assesses the relatedness among these variables. Figure 2 shows the results of the pairwise MI calculations. While high MI is not expected between independently varied growth conditions, a relatively high MI of 0.39 was observed between Ga flux and growth temperature. This correlation arises because Ga flux needs to be adjusted at elevated substrate temperatures to compensate for GaSe evaporation from the surface. Otherwise, low MI here indicates good diversity in training data for our supervised models. Focusing on the correlations between conventional scalar quantities (FHWM and RMS roughness) and growth conditions, we find that for FWHM, the only condition showing a weak correlation (MI = 0.09) was Ga flux, while RMS roughness shows a significant correlation with the Se:Ga flux ratio (MI = 0.5). These MI analysis results align with our previously reported experimental observations,[24,41] which indicate that in MBE growth of GaSe films, crystallinity is primarily influenced by the growth rate, whereas surface morphology is highly sensitive to Se:Ga flux ratio. Interestingly, all three growth conditions exhibit consistent, moderate correlations with the learned RHEED embeddings, with MI of 0.34, 0.30, and 0.28 for the Se:Ga flux ratio, growth temperature, and Ga flux, respectively. On the other hand, the RHEED embeddings show a relatively strong correlation with FWHM (MI = 0.68) and a moderate correlation with RMS roughness (MI = 0.35).

Since the RHEED embeddings contain 2D information, we computed the 1D direction within the embedding space that maximizes MI with the given variable, as illustrated in Figure 3. The direction of maximal MI varies across the variable: Ga flux is aligned primarily along the $Z_1$ axis, while the Se:Ga ratio is rotated by about 60 $^{o}$ along the $Z_1$ axis and the growth temperature is rotated by about 45 $^{o}$ with balanced components along both axes. For scalar quantities, FWHM is predominantly oriented along the $Z_0$ axis, whereas RMS roughness aligns closely with $Z_1$, nearly mirroring the Ga flux direction. This analysis reveals that the 2D RHEED embeddings exhibit moderate MI with all three growth conditions, with each condition occupying distinct 1D directions within the embedding space. The rocking curve FWHM and AFM roughness occupy nearly orthogonal directions within the 2D embedding space. Compared to handcrafted scalar representations of XRD and AFM data, the unsupervised RHEED representation demonstrates stronger correlations with both independent and dependent variables. The RHEED embedding captures distinct aspects of each variable along nearly orthogonal directions, providing physical significance to the $Z_0$ and $Z_1$ axes in the RHEED UMAP (Figure 1).

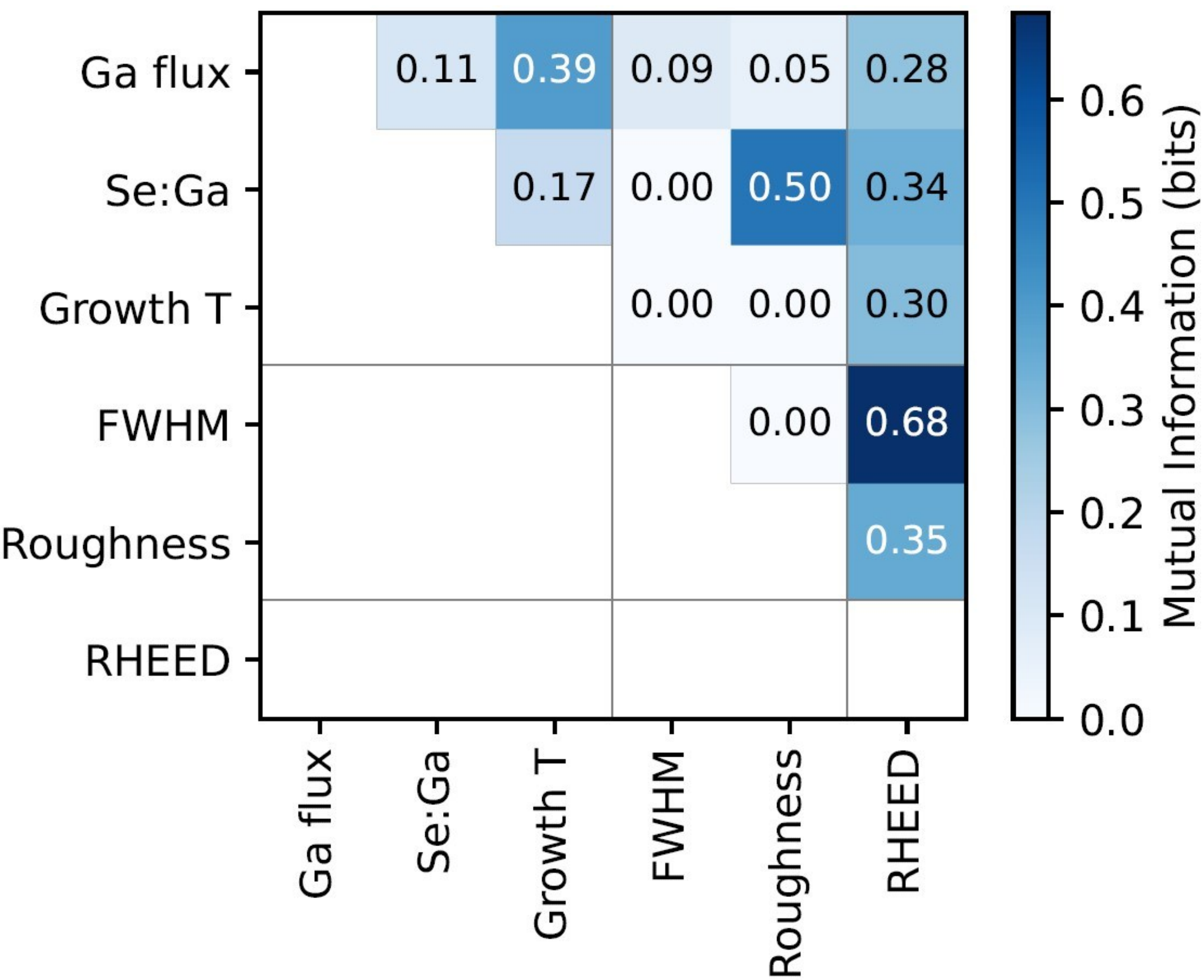

**Figure 2.** Pairwise MI between quantities measured in this study.

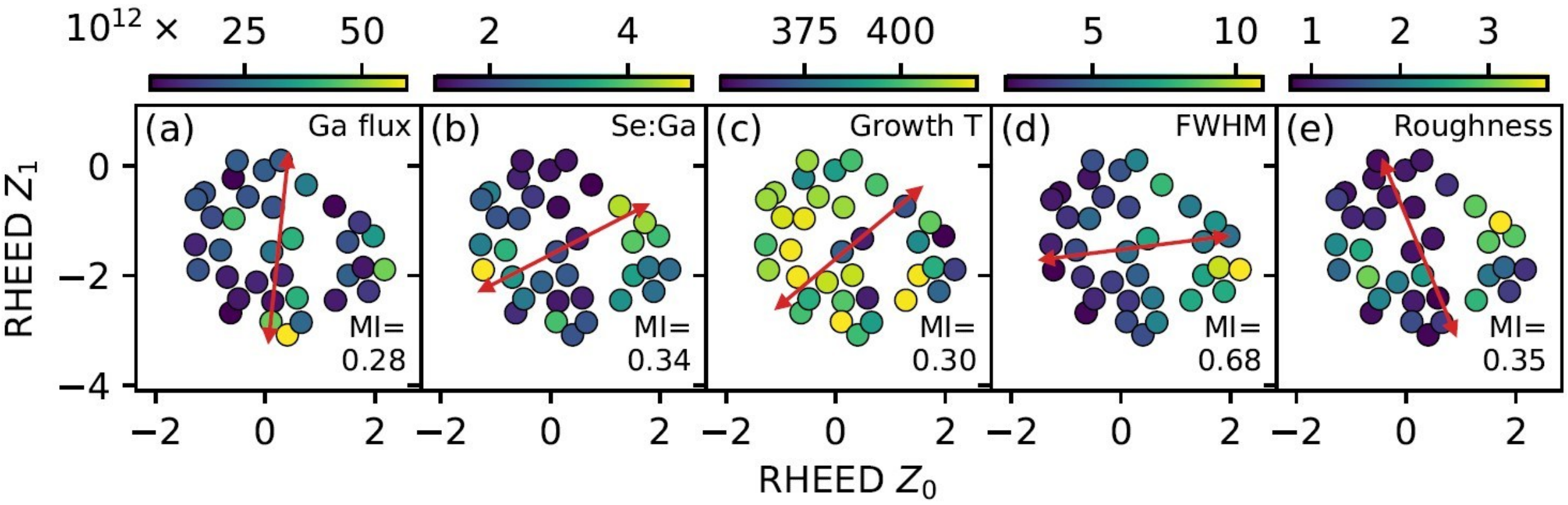

**Figure 3.** All quantities shown in the RHEED embedding space: (a) Ga flux, (b) Se:Ga flux ratio, (c) growth temperature, (d) FWHM of rocking curves, and (e) RMS roughness of AFM images. Red arrows indicate the direction of maximum MI used in Figure 2. The resulting MI is shown in the bottom right of each panel.

**4.3. Supervised Learning.** Having established reliable explainability of the latent features obtained from different characterization modalities and their correlations with meaningful physical features, these latent features can now be used to determine how growth parameters affect the sample properties. For this, we trained regression models to predict the scalar quantities (rocking curve FWHM and AFM RMS roughness) from the growth conditions. The models were further analyzed in the context of anomaly detection. Anomalies are samples exhibiting characteristics that do not align with the typical growth patterns due to random variations or unknown factors, for example, Se flake interference, temperature fluctuations, inaccurate flux measurements, chamber contamination, and unexpected errors with characterizations. Filtering the anomalies out is necessary to provide a high-quality sample database for model construction.

**4.3.1. Rocking Curve FWHM Prediction.** We initiated process modeling using the rocking curve FWHM as the target variable, with growth conditions as input features. To evaluate predictive performance, we employed Random Forest regression models with LOO CV. Model performance was primarily assessed using the coefficient of determination ($R^2$), also known as explained variance. While $R^2$ is typically bounded between 0 and 1 in statistical analysis, its computation on unseen validation data in predictive modeling can yield negative values. A negative $R^2$ indicates that the model introduces additional variance beyond the assumption that all samples are distributed around the mean, meaning its predictions perform worse than random guessing.

Note that while the model performance without the LOO procedure (i.e., performance on only training data) is $R^2 = 0.83$, this does not accurately portray the expected performance on unseen data. Incorporating LOO CV reveals poor predictive performance when all sample data are included, as shown in Figure 4a, where the predicted values exhibit higher variance than simply assuming the mean (i.e., $R^2 < 1$). To address this, we leveraged the RHEED embeddings to refine the dataset by excluding low-quality samples with $Z_0 > 1$. This refinement resulted in a more accurate model (Figure 4b), achieving moderate predictive power with $R^2 = 0.55$. The improved model performance, which significantly surpasses random guessing, demonstrates a stronger alignment with experimental trends, highlighting the effectiveness of RHEED embeddings in enhancing model accuracy for MBE process predictions.

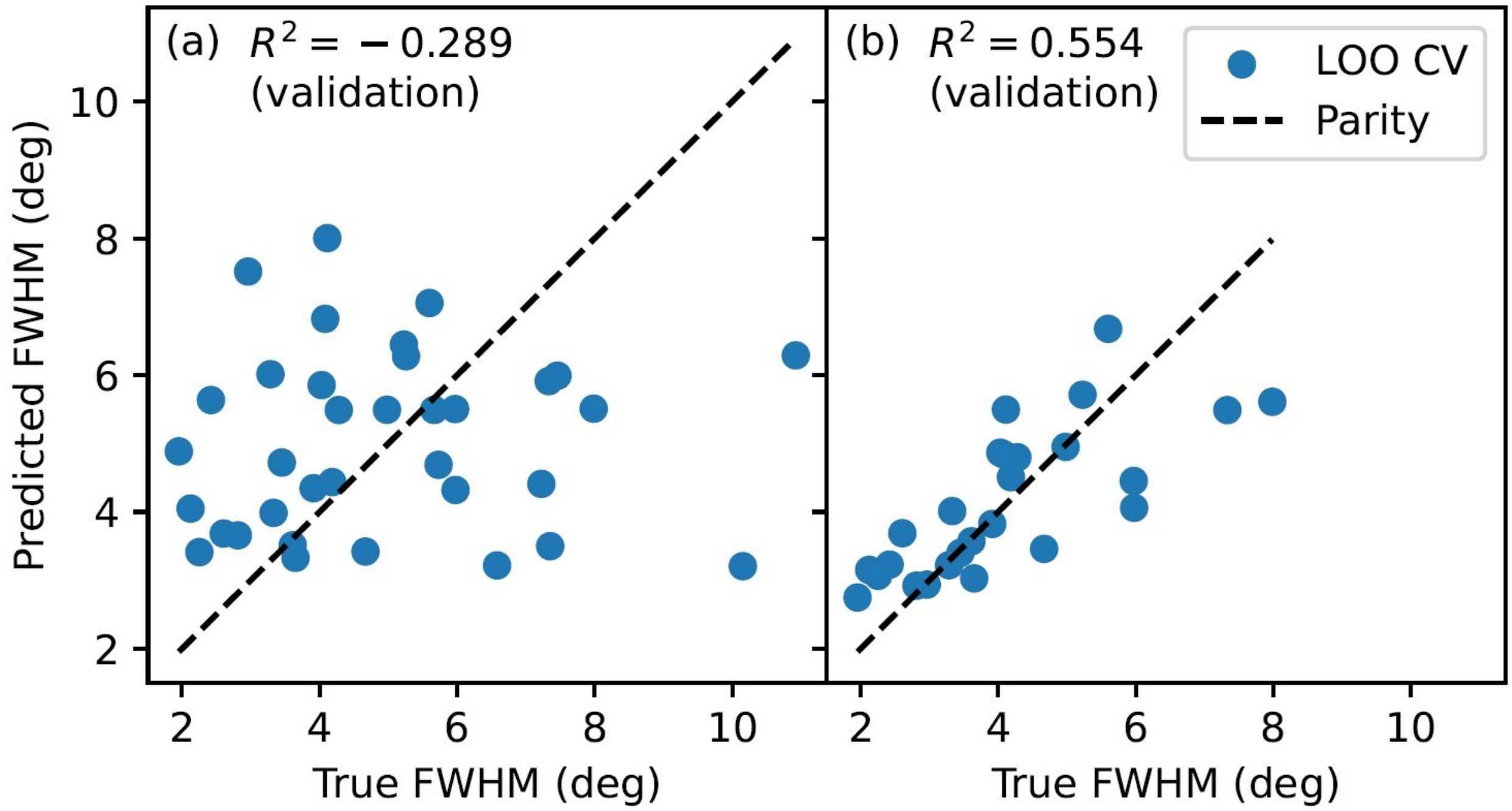

**Figure 4.** Performance of Random Forest Regression models in predicting rocking curve FWHM with LOO CV. Since LOO withholds each sample during training, we refer to every prediction as part of the model validation set. (a) The parity plot when including all samples in the LOO CV. (b) The parity plot when including only samples with RHEED $Z_0 < 1$ in the LOO CV.

**4.3.2. AFM RMS Roughness Prediction.** We repeated the supervised learning procedure for predicting AFM RMS roughness. The model achieves $R^2 = 0.94$ when evaluated on training data without LOO CV, again indicating overfitting and not reflecting its true predictive performance on unseen data. After incorporating LOO CV, the full sample set achieves a predictive capability of $R^2 = 0.56$ (Figure 5a), notably exceeding the predictive accuracy observed for FWHM under similar conditions. Interestingly, excluding samples with $Z_0 > 1$ does not enhance the roughness prediction model; instead, it reduces $R^2$ to 0.47 (Figure 5b). Using the complete dataset, the predictive model for FWHM performed poorly, while the model for RMS roughness showed moderate predictive capability. However, after removing low-quality samples identified by the RHEED embedding, the predictive performance for FWHM improved significantly, whereas the accuracy for roughness prediction slightly declined. For this observation, we propose two possible explanations: First, developing a predictive model for crystallinity depends more heavily on a high-quality sample set than constructing a model for RMS roughness; second, while RHEED is generally considered indicative of both crystallinity and surface morphology, our modeling suggests that crystallinity variations are the primary driver of changes in RHEED. Specifically, the exclusion of low-quality samples based on the RHEED $Z_0$ component primarily targeted samples with abnormal crystallinity, whereas the $Z_1$ component does not exhibit a clear boundary. It is important to emphasize that this does not imply RHEED patterns lack information on surface morphology; rather, establishing a robust correlation between RHEED and RMS roughness remains a challenge.

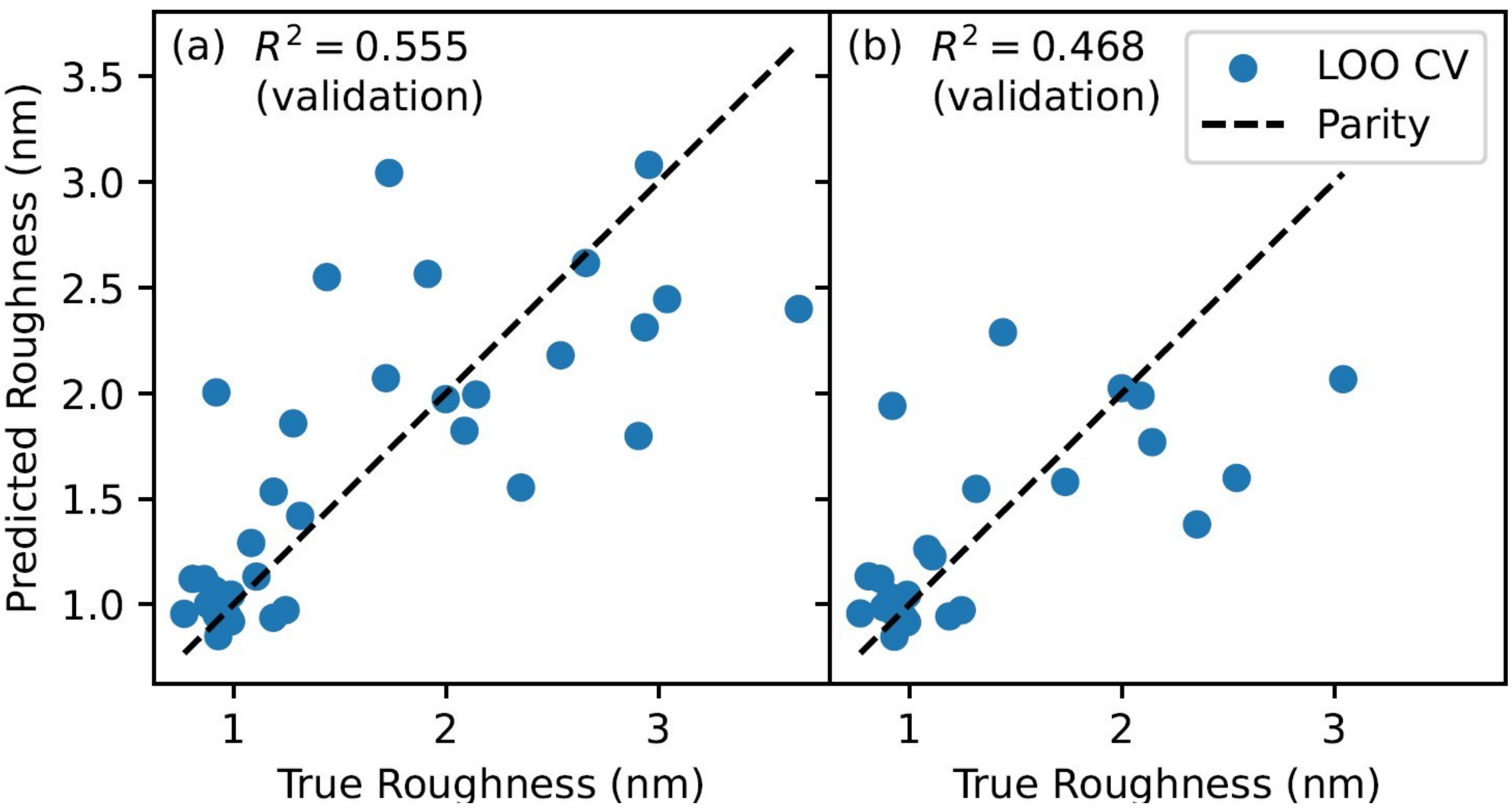

**Figure 5.** Performance of Random Forest regression models in predicting AFM roughness with LOO CV. Since LOO withholds each sample during training, we refer to every prediction as part

of the model validation set. (a) The parity plot when including all samples in the LOO CV. (b) The parity plot when including only samples with $Z_0 < 1$ in the LOO CV.

**4.3.3. Anomaly Detection.** Providing high-quality and self-consistent data is a critical prerequisite for obtaining accurate models; excessively noisy data can only lead to unreliable models that fail to generalize well to unseen inputs. Our preceding analysis demonstrates that RHEED is a promising modality for promptly identifying low-quality samples. Experimental observations further indicate that these low-quality samples are likely outliers, deviating from the typical growth trends established by most experimental data. To validate this, we selected three such samples and repeated the growth using their nominal growth conditions. The resulting samples were noticeably different from the originals, confirming that the low qualities of these samples were not inherent but rather the result of unknown stochastic factors causing anomalous growth behaviors. Details of the experiments, along with comparative RHEED, AFM, and rocking curve FWHM data, are provided in Figures S1-S3. In this section, we compared the anomaly detection capability of the learned RHEED embeddings to another method based solely on measuring discrepancies between similar growth conditions. To this end, we conducted a recursive sample elimination analysis based on the $R^2$ metric computed in LOO CV for predicting rocking curve FWHM. In each round of the procedure, the sample that was least consistent with the rest of the dataset was removed. This was defined as the sample whose removal led to the highest LOO $R^2$. By removing more samples, the model performance generally increased (though not monotonically). The result of this procedure is shown in Figure 6a: four of the first five samples removed were from the $Z_0 > 1$ subset, while the remaining Sample #35 lacked RHEED patterns. After removing the five samples, the model $R^2$ rapidly increased from –0.2 to about 0.3. Subsequently, a few samples with $Z_0 < 1$ were also excluded. Upon removing eleven samples, including seven with $Z_0 > 1$ and three with $Z_0 < 1$, the $R^2$ further improved to around 0.6. Continuing this process only led to a stagnation in performance improvement.

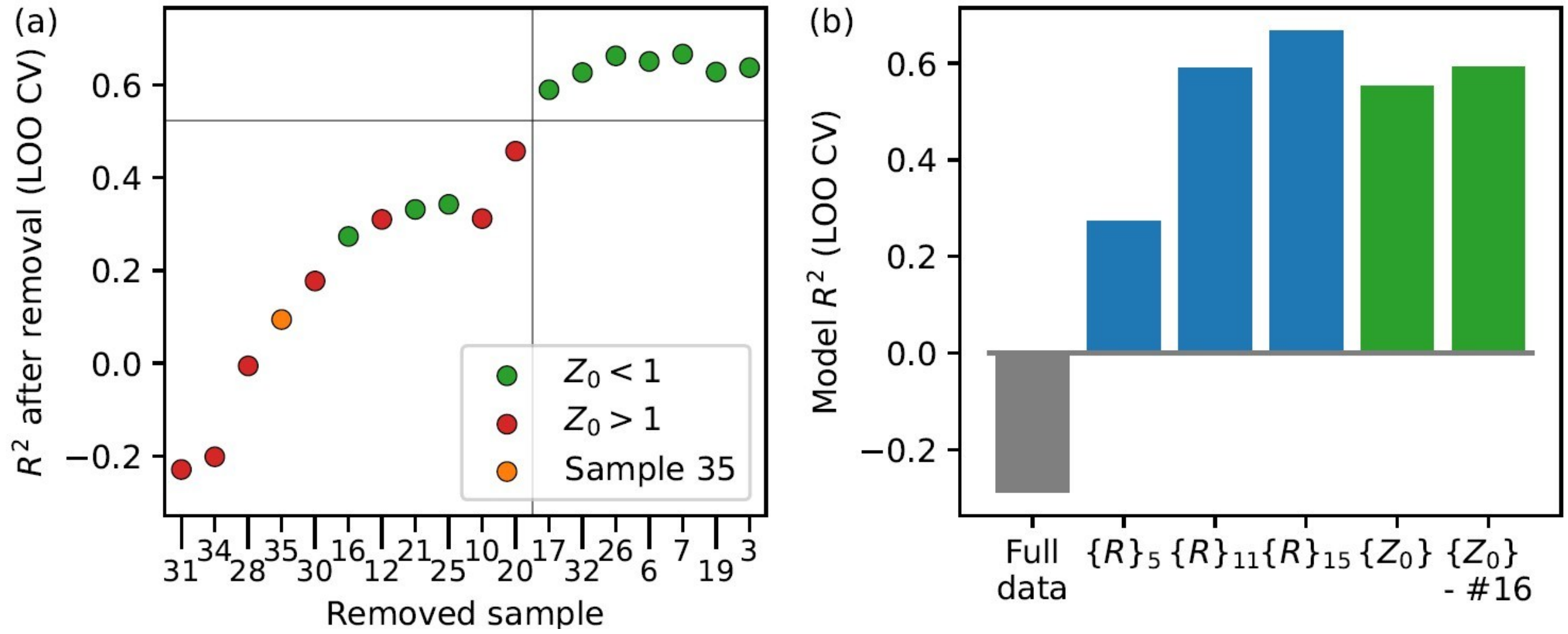

**Figure 6.** (a) Recursive sample elimination based on $R^2$ scores in LOO CV evaluation for predicting rocking curve FWHM. Samples with $Z_0 > 1$ and $Z_0 < 1$ are shown in red and green, respectively. The RHEED pattern of Sample #35 (shown in orange) was lost. (b) Model performance (validation $R^2$) when using different subsets of the data. $\{R\}_N$ designates the set resulting after the $N_{th}$ sample removed by the recursive sample elimination (i.e., panel a) while $\{Z_0\}$ designates the set of only $Z_0 < 1$ samples. $\{Z_0\}$-#16 designates the additional removal of Sample #16, the first non-$\{Z_0\}$ sample to appear in recursive elimination. Colors give a visual grouping of similar sets indicated on the $x$-axis but are not derived from data.

Based on this sample removal procedure, we calculated the LOO $R^2$ for a few different subsets of the data: some different numbers of samples from the recursive elimination procedure, designated with $\{R\}_N$, and some based on the RHEED $Z_0$, designated with $\{Z_0\}$. We show in Figure 6b that the $\{Z_0\}$ set, representing the removal of nine samples with $Z_0 > 1$, offers similar validation performance compared to the recursive removal procedure. An additional calculation was performed on the "$\{Z_0\}$-#16" set, where Sample #16, the first non-$\{Z_0\}$ sample, was excluded in addition to the $Z_0 > 1$ group. This process only slightly improved performance further compared to $\{Z_0\}$. Based on this analysis, we concluded that the RHEED UMAP provided sufficient anomaly capability, and the recursive elimination of samples that had a deleterious effect in LOO CV was not likely to give further improvement to the model performance on unseen data.

Sample #16, the first non-$\{Z_0\}$ sample, has a growth temperature of 419 ºC, the 97$^{th}$ percentile of that parameter, which had a maximum of 420 ºC (Samples #8 and #28 were also grown at 419 ºC and were excluded by the RHEED $Z_0$ filter, though Samples #18 and #37 were not). This may be the reason that it was selected for removal in the recursive procedure. However, removing it only resulted in a slight increase in model performance compared to $Z_0$ alone and did not lead to qualitatively superior predictions.

**4.4.4. Feature Importance.** We next evaluated the feature importance within the models trained exclusively on the $\{Z_0\}$ set using drop-column feature importance, where the model is retrained in the absence of the dropped feature. Bootstrap sampling with replacement was used to estimate a standard error. When an important feature (or "column") is dropped from the feature space, we expect a significant decline in model performance, which confirms that accurate predictions of the target variable depend on access to that feature. Conversely, the removal of unimportant or redundant features leads to negligible change in performance because it is either irrelevant to the target variable or its information content can be obtained through analyzing another feature.

The results for drop-column feature importance on rocking curve FWHM and RMS roughness predictive models are presented in Figure 7. FWHM exhibits a strong dependence on Ga flux (as the explained variance drops nearly to zero when Ga flux is removed), and a weaker dependence on the Se:Ga ratio, while growth temperature has a negligible effect. In contrast, RMS roughness depends somewhat on the Se:Ga ratio (as the explained variance approximately halving when Se:Ga ratio is dropped), followed by a weaker dependence on Ga flux and minimal impact from growth temperature. These findings align with previous MI analysis and are consistent with experimental observations.[24,41] Our experiments[24] indicate that within the temperature window suitable for GaSe crystal growth, variations in temperature do not significantly alter crystallinity. Outside of this range, however, amorphous samples are formed. Therefore, we believe that, for GaSe growth, while growth temperature plays a crucial role in determining what kind of phase is formed, it does not significantly influence the crystallinity of the crystalline samples. This behavior may be attributed to the relatively low sublimation temperature of GaSe (i.e., about 500 °C),[74] where higher temperatures tend to cause material loss rather than promote further crystallization. Additionally, the interaction between the GaAs substrate and GaSe is relatively strong, reducing the sensitivity of crystallinity to substrate temperature in the early growth stages. As for surface morphology, we experimentally observed some dependence on substrate temperature; however, this effect is less significant and not as sensitive as the influence of the flux ratio. Moreover, the impact of growth temperature on surface morphology appears to be coupled with Ga flux, as confirmed by both experimental observations and MI analysis. This interdependence reduces the independent predictive importance of growth temperature on surface roughness.

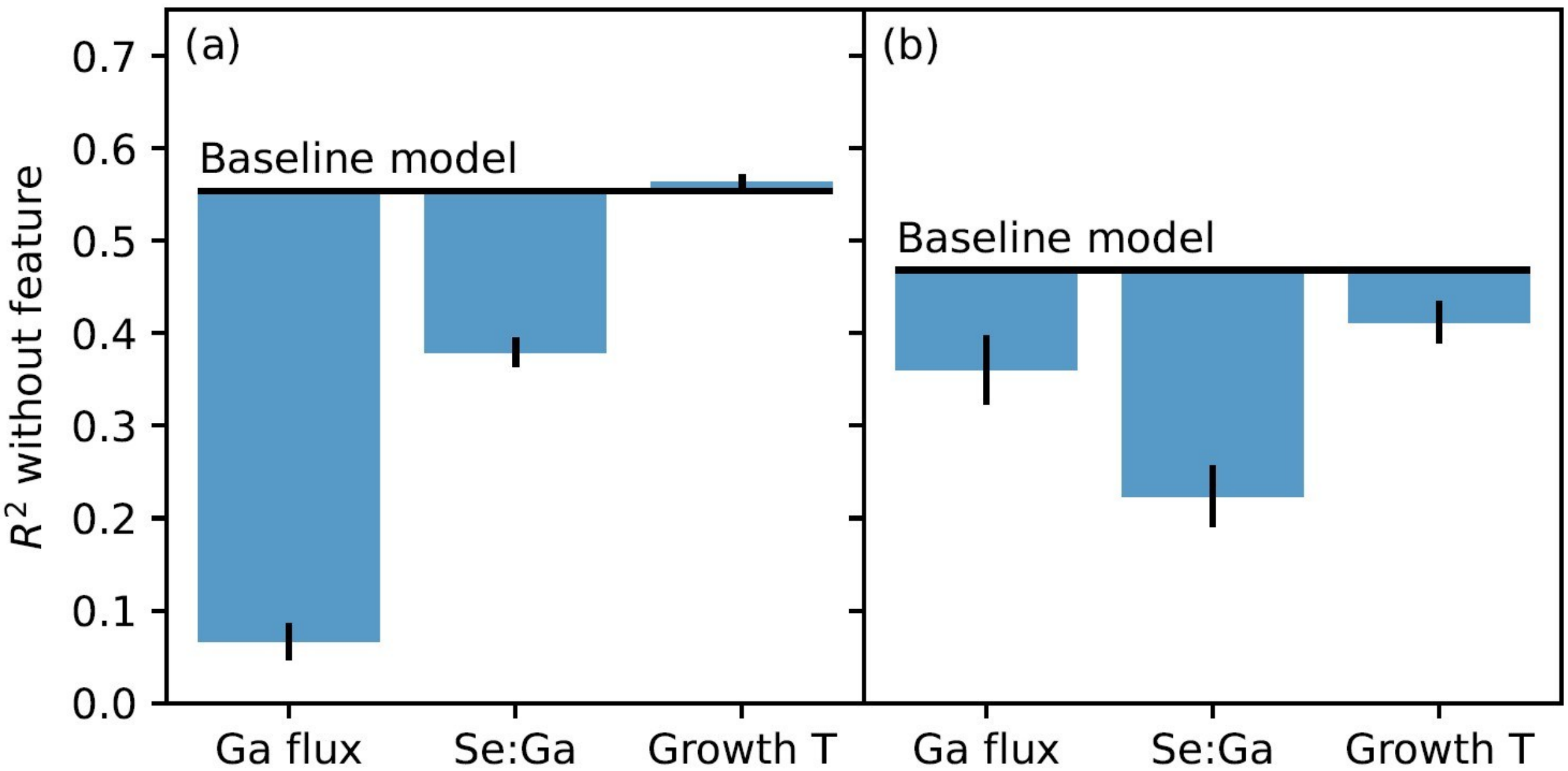

**Figure 7.** Drop-column feature importance for (a) rocking curve FWHM and (b) RMS roughness for Random Forest surrogate models. Performance evaluated with LOO CV. Error bars are the standard error from 100 bootstrapped samples with replacement.

Incorporating the RHEED embeddings in the feature set reveals that RHEED provides independent information compared to the Ga flux or Se:Ga ratio alone. As shown in Figure 8, the baseline performance improves slightly compared to Figure 7, especially for FWHM, while feature importance also changes drastically. The FWHM model, instead of relying primarily on Ga flux and Se:Ga ratio, now substitutes RHEED $Z_0$ for Se:Ga as the second most important feature. Although other features contribute, their influence is considerably weaker. Note that the FWHM model with neither Ga flux nor RHEED $Z_0$ yields $R^2 = -0.05$, indicating that at least one of those two is necessary for accurate predictions. For the AFM roughness model, adding RHEED $Z_1$ did not significantly enhance overall performance. However, it took some precedence away from the growth conditions, with a less deleterious effect when Ga flux was removed. As in the case of the FWHM model, the roughness model without Se:Ga or RHEED $Z_1$ gives a lower performance of $R^2 = 0.24$, which matches the performance reported in Figure 7b. The improvement suggests that RHEED measurements provide additional information relevant to FWHM prediction.

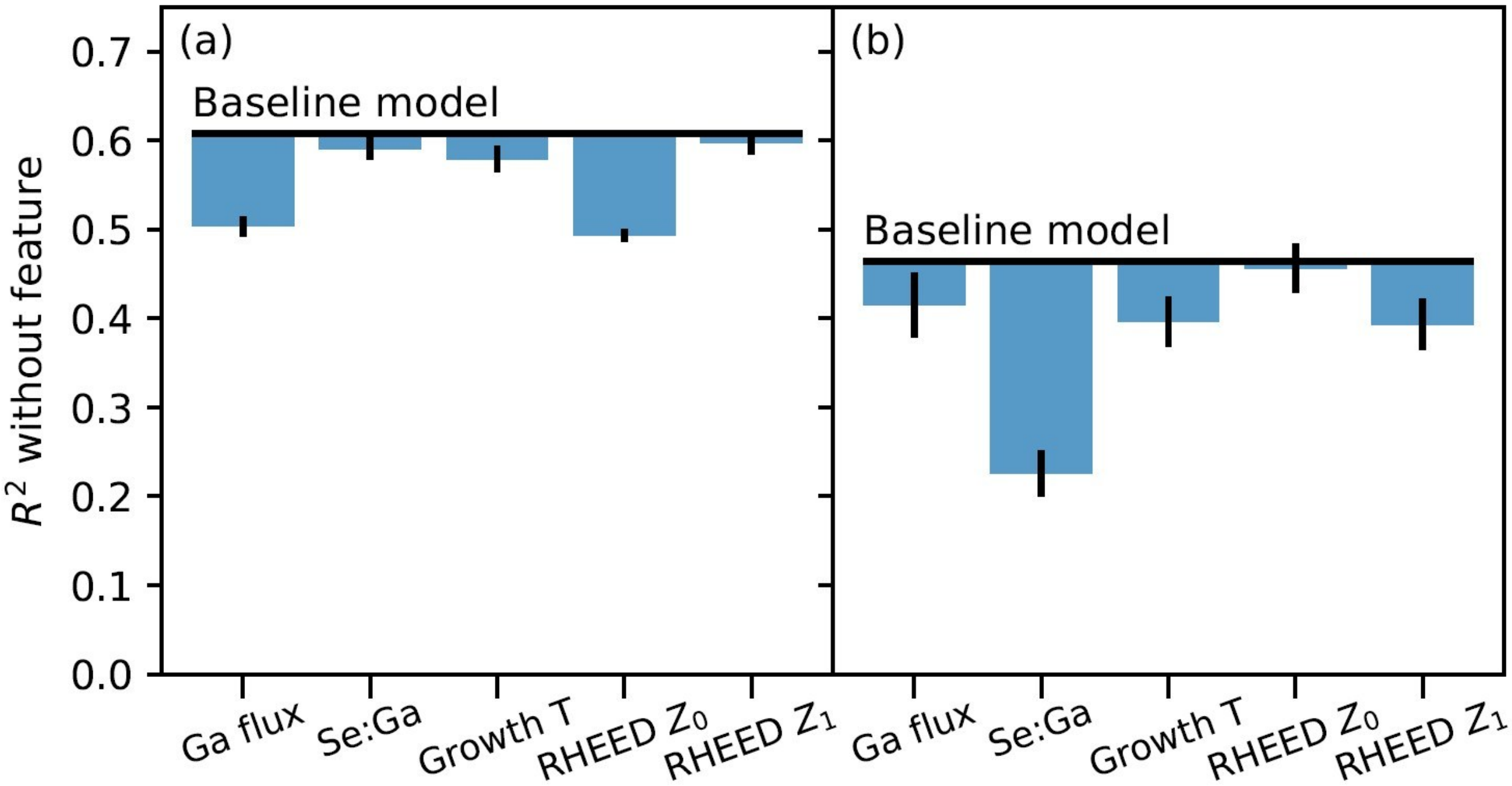

**Figure 8.** Drop-column feature importance for (a) rocking curve FWHM and (b) RMS roughness for Random Forest surrogate models, including learned RHEED embeddings as features. Performance evaluated with LOO CV. Error bars are the standard error from 100 bootstrapped samples with replacement.

## 5. CONCLUSIONS

The effectiveness of unsupervised learning on RHEED UMAP embeddings has been demonstrated in distinguishing between high- and low-quality samples with a well-defined boundary. Our approach successfully captures physically meaningful image features in a manner comparable to human experts. The learned RHEED embeddings exhibit a strong correlation with the crystallinity parameter FWHM obtained from ω scans but only a weak correlation with surface RMS roughness.

Additionally, MI analysis reveals that the RHEED embeddings retain more information about growth conditions than traditional scalar characterization metrics.

Among the three key growth parameters, growth rate has the greatest impact on FWHM, while the Se:Ga flux ratio primarily affects RMS roughness and the RHEED $Z_0$ component. Growth temperature has a negligible effect on FWHM but slightly influences roughness. Our process models show predictive capability for conventional scalar sample features, including rocking curve FWHM and AFM RMS roughness. Incorporating *in-situ* RHEED embeddings significantly improves model performance in predicting rocking curve FWHM. Furthermore, anomaly detection via residual analysis in supervised learning closely aligns with unsupervised learning from *in-situ* measurements, reinforcing the reliability of the FWHM predictive model and cross-validation strategy.

While the RHEED signals used in our analysis were acquired at the end of sample growth, their rapid analysis can accelerate iterative process optimization by avoiding the need for *ex situ* characterization. The high degree of MI between *in situ* RHEED and *ex situ* XRD and AFM also show that RHEED can be used to corroborate those results. Thus, our findings highlight the potential of machine learning-assisted *in situ* RHEED analysis for enhancing real-time monitoring and control in MBE.

ASSOCIATED CONTENT

**Supporting Information**

The table listing the sample numbers and growth conditions; the table listing the machine learning-derived characterization data (FWHM, RMS roughness, and RHEED embedding component $Z$); RHEED patterns, AFM images, and rocking curves of Samples #8 and #16 (control group 1);

RHEED patterns, AFM images, and rocking curves of Samples #28 and #18 (control group 2); RHEED patterns, AFM images, and rocking curves of Samples #34 and #13 (control group 3).

AUTHOR INFORMATION

**Corresponding Author**

*Stephanie Law. E-mail: slaw@psu.edu.

**Author Contributions**

The manuscript was written through the contributions of all authors. All authors have given approval to the final version of the manuscript.

**Funding Sources**

NSF Cooperative Agreement DMR-2039351.

**Notes**

The authors declare no competing financial interest.

ACKNOWLEDGMENT

This study is based upon research conducted at the Pennsylvania State University Two-Dimensional Crystal Consortium—Materials Innovation Platform (2DCC-MIP), which is supported by NSF Cooperative Agreement DMR-2039351.

ABBREVIATIONS

MBE, molecular beam epitaxy; RHEED, reflection high-energy electron diffraction; XRD, X-ray diffraction; AFM, atomic force microscopy; FWHM, full-width at half maximum; RMS, root-mean-square; 2D, two-dimensional; vdW, van der Waals; MI, mutual information; UMAP, uniform manifold approximation and projection; KSG, Kraskov-Stögbauer-Grassberger; LOO, Leave-One-Out; CV, cross-validation; $R^2$, coefficient of determination.

Table of Contents (TOC) Figure

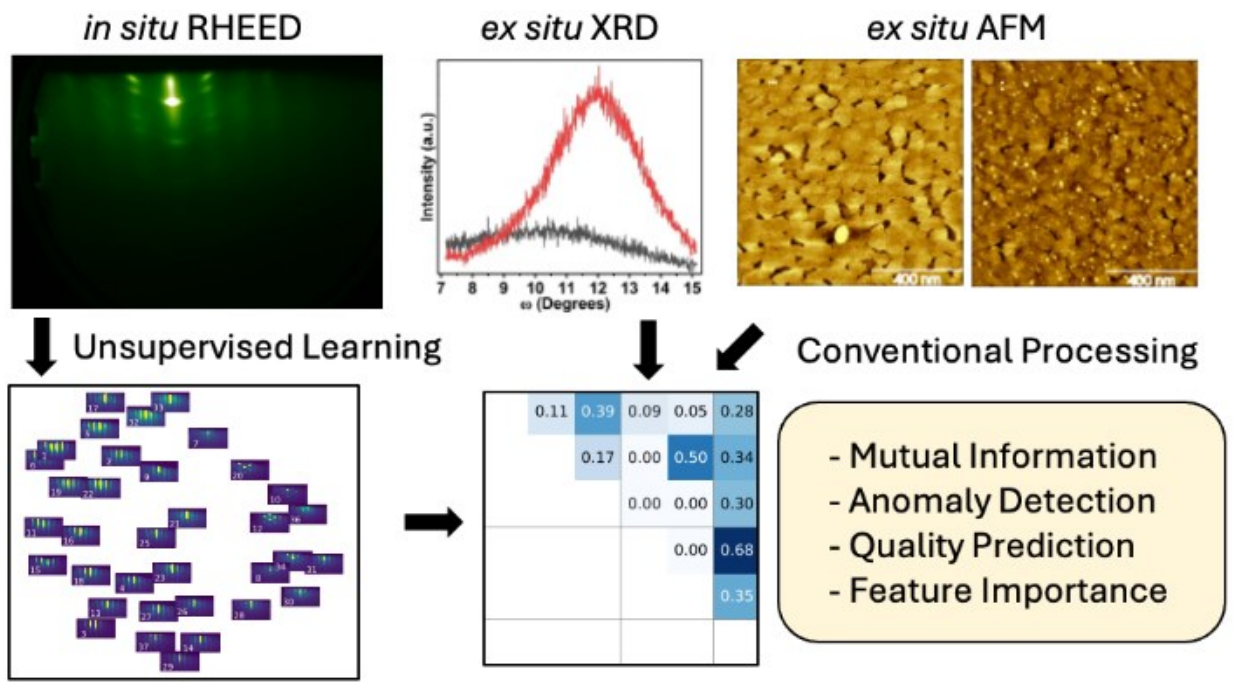

**Supporting Information**

# Multi-modal machine learning analysis of GaSe molecular beam epitaxy growth conditions

*Mingyu Yu,* [†] *Isaiah A. Moses,* [‡] *Wesley F. Reinhart,* [§, ∥] *Stephanie Law*[*,§]

[†]Department of Materials Science and Engineering, University of Delaware, Newark, Delaware 19716, United States

[‡]Materials Research Institute, Pennsylvania State University, University Park, Pennsylvania 16827, United States

[§]Department of Materials Science and Engineering, Pennsylvania State University, University Park, Pennsylvania 16827, United States

[∥]Institute for Computational and Data Science, Pennsylvania State University, University Park, Pennsylvania 16827, United States

[*]Corresponding author: Stephanie Law

E-mail: slaw@psu.edu

**Table S1.** Growth conditions of 37 GaSe film samples grown on GaAs (111)B substrates. Sample # is the sample number used in the main text. LiST # is the original sample number in the LiST database. Fluxes were measured by quartz crystal microbalance. Growth temperatures were measured by a thermocouple. Raw data on LiST can be found in Ref. 1.

| Sample # | LiST # | Ga flux [$\times 10^{13}$ atoms cm$^{-2}$ s$^{-1}$] | Se:Ga flux ratio | Growth temp. [ºC] |
|---|---|---|---|---|
| 1 | 240221B | 1.9100 ± 0.0100 | 2.825 ± 0.015 | 410 |
| 2 | 240309A | 1.9500 ± 0.0100 | 1.840 ± 0.010 | 410 |
| 3 | 240405A | 0.4975 ± 0.0095 | 1.705 ± 0.035 | 400 |
| 4 | 240604A | 1.0750 ± 0.0050 | 2.400 ± 0.020 | 414 |
| 5 | 240405B | 0.5380 ± 0.0100 | 1.580 ± 0.030 | 385 |
| 6 | 240306A | 2.0100 ± 0.0100 | 2.330 ± 0.010 | 410 |
| 7 | 240430A | 2.8300 ± 0.0100 | 1.220 | 401 |
| 8 | 240525C | 2.1900 ± 0.0100 | 3.335 ± 0.015 | 419 |
| 9 | 240221A | 1.9100 ± 0.0100 | 1.310 ± 0.010 | 410 |
| 10 | 240617B | 1.4500 ± 0.0100 | 4.380 ± 0.030 | 404 |
| 11 | 240306B | 1.0600 ± 0.0100 | 2.795 ± 0.025 | 400 |
| 12 | 240616B | 1.8900 ± 0.0100 | 3.965 ± 0.015 | 385 |
| 13 | 240608B | 0.8575 ± 0.0095 | 2.730 ± 0.030 | 395 |
| 14 | 240426A | 2.3300 ± 0.0100 | 2.215 ± 0.005 | 390 |
| 15 | 240221C | 1.9100 ± 0.0100 | 4.920 ± 0.020 | 410 |
| 16 | 240604B | 2.1300 ± 0.0100 | 3.605 ± 0.015 | 419 |
| 17 | 240402B | 2.1100 ± 0.0100 | 1.435 ± 0.005 | 410 |
| 18 | 240608C | 0.9765 ± 0.0095 | 3.195 ± 0.035 | 419 |
| 19 | 240424A | 1.7500 | 2.180 | 416 |
| 20 | 240616A | 0.5775 ± 0.0095 | 4.505 ± 0.075 | 371 |
| 21 | 240430B | 4.0050 ± 0.0050 | 1.535 ± 0.005 | 359 |
| 22 | 240501B | 4.3450 ± 0.0050 | 2.175 ± 0.005 | 418 |
| 23 | 240525B | 1.1350 ± 0.0050 | 2.275 ± 0.015 | 414 |
| 24 | 240314C | 1.4500 | 5.200 | 383 |
| 25 | 240501A | 2.6500 ± 0.0100 | 2.075 ± 0.005 | 374 |
| 26 | 240604C | 3.9850 ± 0.0050 | 1.565 ± 0.005 | 359 |
| 27 | 240608A | 0.7770 ± 0.0100 | 1.470 ± 0.020 | 402 |
| 28 | 240525A | 0.9370 ± 0.0090 | 3.090 ± 0.030 | 419 |
| 29 | 240426B | 5.9550 ± 0.0050 | 2.180 | 399 |
| 30 | 240617A | 1.3950 ± 0.0050 | 2.540 ± 0.020 | 375 |
| 31 | 240424B | 4.8000 | 2.510 | 366 |
| 32 | 240402C | 2.1100 ± 0.0100 | 1.435 ± 0.005 | 390 |
| 33 | 240402D | 2.1300 ± 0.0100 | 1.420 ± 0.010 | 400 |
| 34 | 240521B | 0.8575 ± 0.0095 | 2.690 ± 0.030 | 395 |
| 35 | 240521A | 0.7970 ± 0.0090 | 1.530 ± 0.020 | 402 |
| 36 | 240422A | 3.4250 ± 0.0050 | 3.770 ± 0.010 | 353 |
| 37 | 240422B | 4.7000 ± 0.0100 | 3.825 ± 0.005 | 420 |

**Table S2.** Characterization results of 37 GaSe film samples. Sample # and LiST # are the sample numbers used in the main text and the LiST database, respectively. FWHM refers to the full-width at half maximum of a rocking curve obtained from high-resolution X-ray diffraction ω scan. The RMS (root-mean-square) roughness was calculated from atomic force microscopy (AFM) scans over a 1 cm × 1 cm area. RHEED $Z_0$ and $Z_1$ are the components of reflection high-energy electron diffraction embeddings and they have no physical units. RHEED patterns of Samples #24 and #35 were lost due to technical issues. Raw data on LiST can be found in Ref. 1. Processed data and code can be found in Ref. 2. Uncertainty in FWHM and RMS roughness are estimated by the mean absolute deviation when repeating analysis of the same samples.

| Sample # | LiST # | FWHM [± 0.086°] | RMS roughness [± 0.067 nm] | RHEED $Z_0$ [a.u.] | RHEED $Z_1$ [a.u.] |
|---|---|---|---|---|---|
| 1 | 240221B | 2.25 | 0.92 | -1.11 | -0.49 |
| 2 | 240309A | 3.65 | 0.99 | -0.32 | -0.57 |
| 3 | 240405A | 2.12 | 0.86 | -0.63 | -2.68 |
| 4 | 240604A | 3.61 | 2.00 | -0.16 | -2.12 |
| 5 | 240405B | 2.42 | 0.92 | -0.58 | -0.23 |
| 6 | 240306A | 2.61 | 1.31 | -1.26 | -0.62 |
| 7 | 240430A | 7.99 | 1.19 | 0.75 | -0.35 |
| 8 | 240525C | 7.23 | 1.91 | 1.51 | -2.0 |
| 9 | 240221A | 3.91 | 0.91 | 0.14 | -0.76 |
| 10 | 240617B | 6.58 | 3.66 | 1.72 | -1.03 |
| 11 | 240306B | 2.81 | 2.14 | -1.28 | -1.44 |
| 12 | 240616B | 5.73 | 2.94 | 1.5 | -1.39 |
| 13 | 240608B | 2.96 | 2.09 | -0.49 | -2.43 |
| 14 | 240426A | 5.97 | 1.25 | 0.65 | -2.86 |
| 15 | 240221C | 1.95 | 1.73 | -1.23 | -1.90 |
| 16 | 240604B | 4.03 | 2.54 | -0.82 | -1.54 |
| 17 | 240402B | 4.18 | 0.88 | -0.52 | 0.09 |
| 18 | 240608C | 3.29 | 3.04 | -0.69 | -2.03 |
| 19 | 240424A | 3.33 | 1.08 | -0.97 | -0.94 |
| 20 | 240616A | 5.67 | 2.96 | 1.26 | -0.74 |
| 21 | 240430B | 7.33 | 0.93 | 0.49 | -1.33 |
| 22 | 240501B | 4.98 | 1.11 | -0.57 | -0.96 |
| 23 | 240525B | 4.67 | 2.35 | 0.30 | -2.00 |
| 24 | 240314C | 3.65 | 3.06 | -- | -- |
| 25 | 240501A | 5.23 | 0.92 | 0.12 | -1.57 |
| 26 | 240604C | 5.60 | 0.77 | 0.58 | -2.41 |
| 27 | 240608A | 3.45 | 0.99 | 0.14 | -2.47 |
| 28 | 240525A | 7.35 | 2.66 | 1.28 | -2.45 |
| 29 | 240426B | 4.11 | 0.81 | 0.40 | -3.08 |
| 30 | 240617A | 7.46 | 1.28 | 1.88 | -2.29 |
| 31 | 240424B | 10.92 | 1.19 | 2.17 | -1.89 |
| 32 | 240402C | 4.28 | 0.97 | -0.02 | -0.08 |
| 33 | 240402D | 5.98 | 0.96 | 0.29 | 0.10 |
| 34 | 240521B | 10.15 | 1.72 | 1.79 | -1.85 |
| 35 | 240521A | 5.81 | 1.35 | -- | -- |
| 36 | 240422A | 5.26 | 2.91 | 1.97 | -1.28 |
| 37 | 240422B | 4.08 | 1.44 | 0.10 | -2.84 |

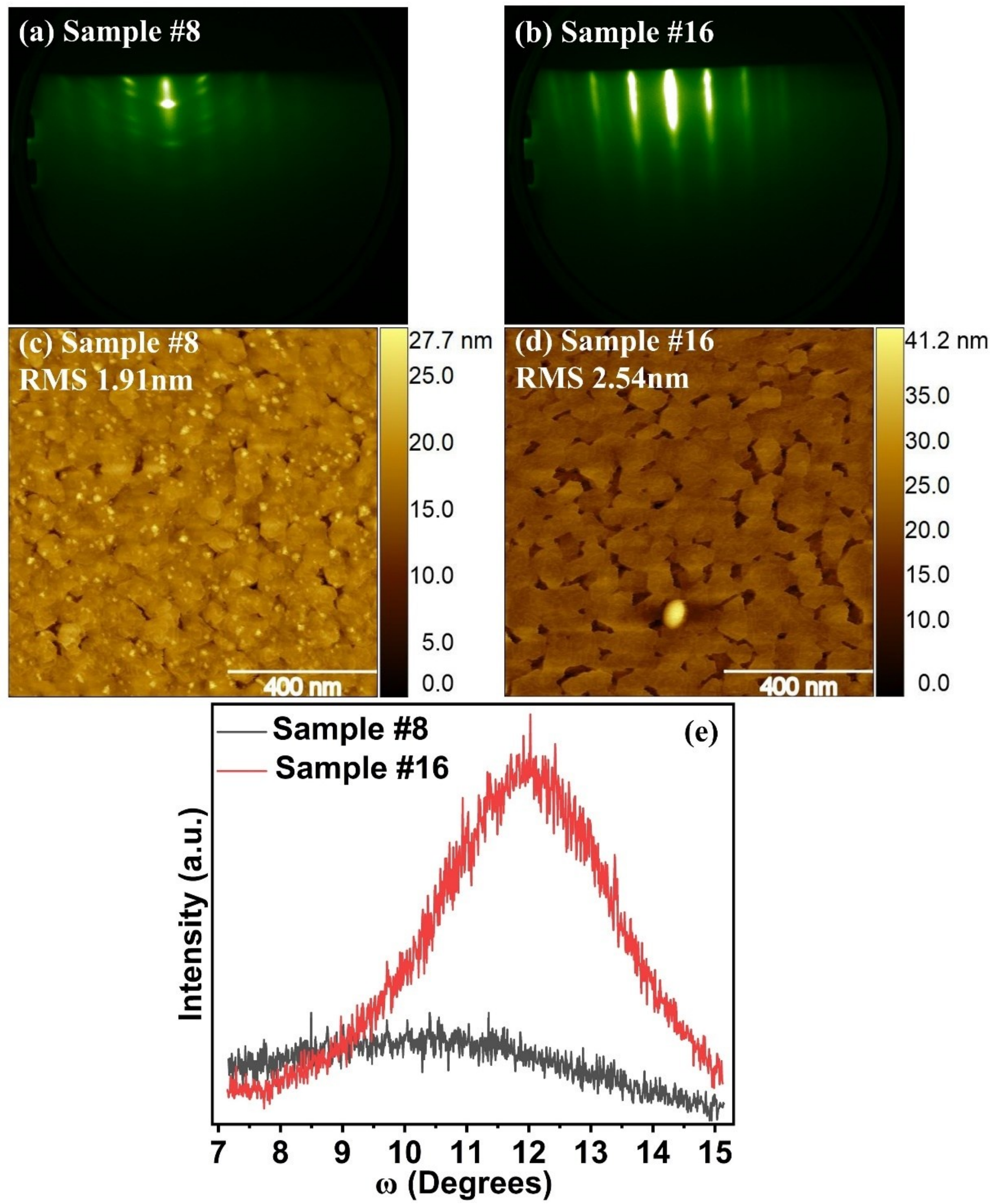

**Figure S1.** Control group 1: Sample #8 (anomalous) and Sample #16 (normal). RHEED patterns recorded along the GaSe $[1\bar{1}00]$ direction for Sample #8 (a) and Sample #16 (b). AFM images of Sample #8 (c) and Sample #16 (d). Rocking curves of Samples #8 and #16 (e). Both samples were grown under the same nominal conditions but on different dates (see Table S1 for details).

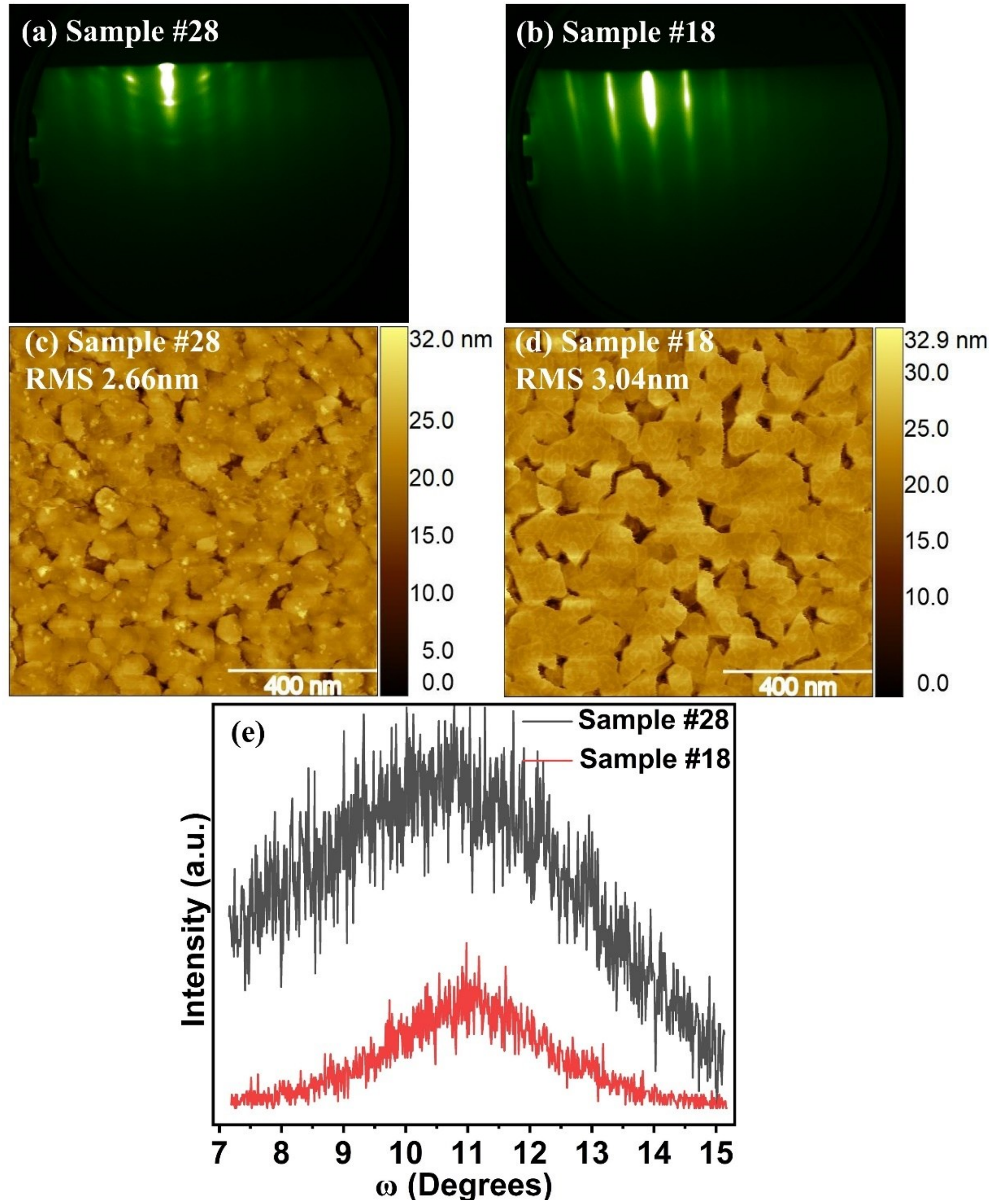

**Figure S2.** Control group 2: Sample #28 (anomalous) and Sample #18 (normal). RHEED patterns recorded along the GaSe $[1\bar{1}00]$ direction for Sample #28 (a) and Sample #18 (b). AFM images of Sample #28 (c) and Sample #18 (d). Rocking curves of Samples #28 and #18 (e). Both samples were grown under the same nominal conditions but on different dates (see Table S1 for details).

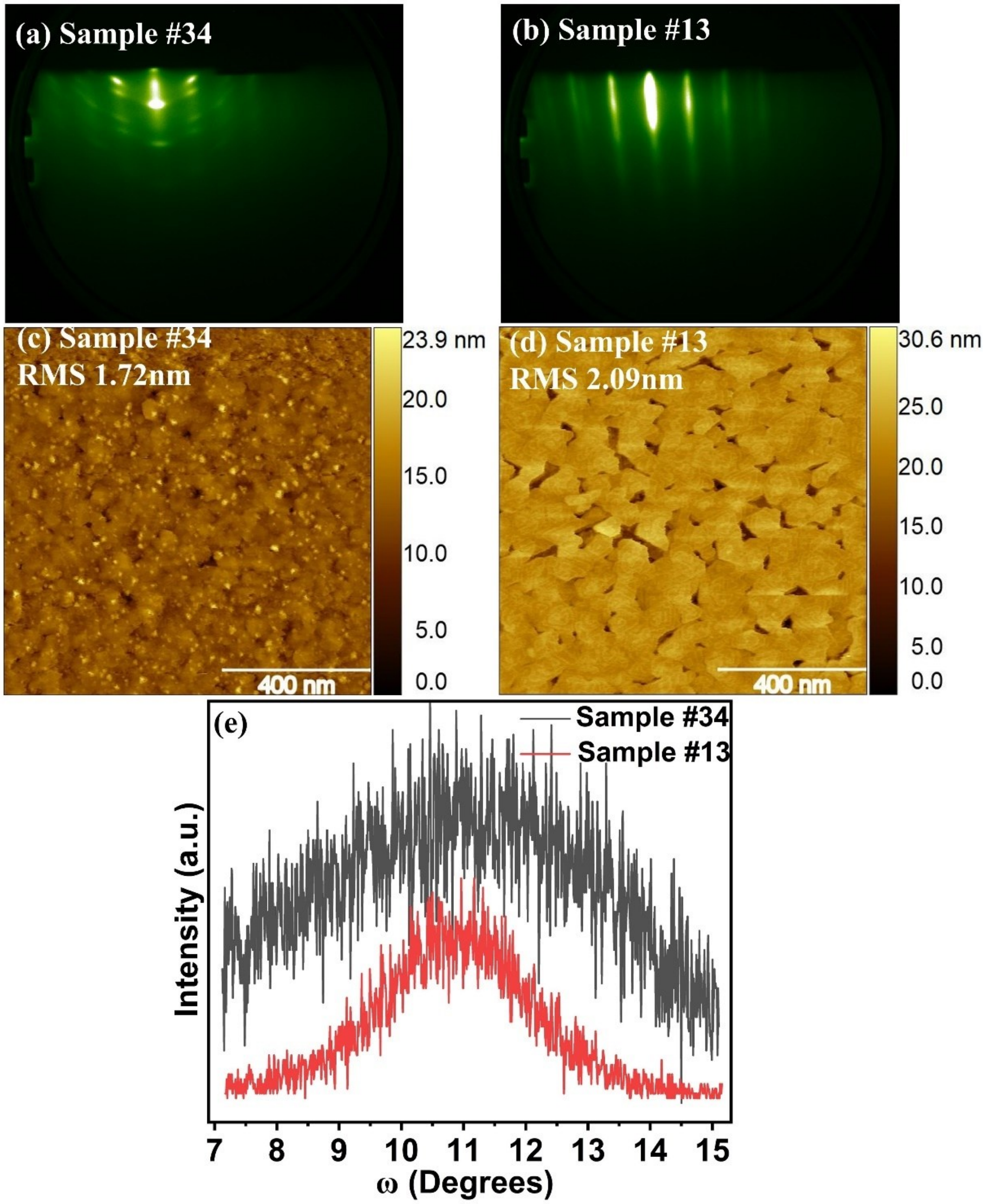

**Figure S3.** Control group 3: Sample #34 (anomalous) and Sample #13 (normal). RHEED patterns recorded along the GaSe $[1\bar{1}00]$ direction for Sample #34 (a) and Sample #13 (b). AFM images of Sample #34 (c) and Sample #13 (d). Rocking curves of Samples #34 and #13 (e). Both samples were grown under the same nominal conditions but on different dates (see Table S1 for details).